 \documentclass[sigconf]{acmart}

\usepackage{caption}
\usepackage{subcaption}
\AtBeginDocument{%
  \providecommand\BibTeX{{%
    \normalfont B\kern-0.5em{\scshape i\kern-0.25em b}\kern-0.8em\TeX}}}

\setcopyright{acmcopyright}
\copyrightyear{2021}
\acmYear{2021}
\acmDOI{10.1145/nnnnn.n}

\acmConference[DSHealth '21]{ Proceedings of DSHealth '21: KDD Workshop on Applied Data Science for Healthcare}{August 14--18, 2021}{DSHealth, Virtual}
\acmBooktitle{ Proceedings of DSHealth '21: KDD Workshop on Applied Data
Science for Healthcare,
  August 14--18, 2021, DSHealth, Virtual}
\acmPrice{15.00}
\acmISBN{978-1-4503-XXXX-X/18/06}

\usepackage[export]{adjustbox}
\begin{document}

\title{Towards a fairer reimbursement system for burn patients using cost-sensitive classification}


\author{Chimdimma Noelyn Onah*\\
Richard Allmendinger\\
Julia Handl}
\affiliation{%
  \institution{University of Manchester, UK}
   \streetaddress{Oxford Road}
   \city{Manchester}
   \country{United Kingdom}
 }
\email{chimdimma.onah@postgrad.manchester.ac.uk}

\author{Ken W. Dunn}
\affiliation{%
  \institution{University Hospital South Manchester, UK}
  \streetaddress{Southmoor Road}
  \city{Wythenshawe}
  \country{United Kingdom}
  }
\email{ken.dunn@mft.nhs.uk}

\renewcommand{\shortauthors}{C.N. Onah, R. Allmendinger, J. Handl, and K.W. Dunn}

\begin{abstract}
The adoption of the Prospective Payment System (PPS) in the UK National Health Service (NHS) has led to the creation of patient groups called Health Resource Groups (HRG). HRGs aim to identify groups of clinically similar patients that share similar resource usage for reimbursement purposes.  These groups are predominantly identified based on expert advice, with homogeneity checked using the length of stay (LOS). However, for complex patients such as those encountered in burn care, LOS is not a perfect proxy of resource usage, leading to incomplete homogeneity checks. To improve homogeneity in resource usage and severity, we propose a data-driven model and the inclusion of patient-level costing. We investigate whether a data-driven approach that considers additional measures of resource usage can lead to a more comprehensive model. In particular, a cost-sensitive decision tree model is adopted to identify features of importance and rules that allow for a focused segmentation on resource usage (LOS and patient-level cost) and clinical similarity (severity of burn). The proposed approach identified groups with increased homogeneity compared to the current HRG groups, allowing for a more equitable reimbursement of hospital care costs if adopted. 
\end{abstract}



\maketitle

\section{Introduction}
The National Health Service (NHS) serves a broad UK population with varied demographic and medical histories. An acceleration in medical advances has led to a greater scope of treating chronic illness~\cite{Naylor2015TransformingCommissioners}. In turn, this burden the NHS with rising costs and with limitation of access to funds, these have fuelled the popularity and the adoption of Prospective Payment System (PPS) over retrospective systems. PPS is a reimbursement method that allows for the predetermination of a fixed amount to be allocated to health providers for the treatment of patients with a specific diagnosis. The overall aim is to ensure an equitable and fair distribution of funds. It also acts as an instrument to understand the organisational activity, such as the type of patient cared for and treatment delivered. The system adopted in the UK is called Health Resource Groups (HRGs). HRGs are groupings of clinically similar patients which use comparable healthcare resources.  Although these groups are generated using patient-level data, the grouping rules are typically generated by transcribing expert advice into if-else rules meant to capture differing patient severity and length of stay (LOS). 

These HRGs should be clinically meaningful groups of diagnoses and interventions that consume similar levels of NHS resources. However, the aim of having homogeneous groups in terms of resource usage might not be met due to: (i) The methodology’s dependence on expert advice, which carries the risk of ignoring less established factors that account for specific patient subgroups’ complexity; (ii) the validation of identified groups using LOS only, an imperfect indicator of resource usage~\cite{Street2012}; (iii) the determination of payment rate as an average of the cost of each HRG. This is, in essence, a generalisation that will exacerbate the effect of any non-homogeneous groups. Non-homogeneity in the resource usage of the identified groups may lead to a significant disadvantage to small and highly specialised services that, by necessity, operate at very high expenditure. Burn services are one such service -- they are small, rely on specialist equipment and intervention, deal with various case complexity and stay open regardless of the number of patients admitted, with a minimum staff number constantly on the rota \cite{NHSEngland2018NHSCasualties}. 

We investigate if the current burn HRG can be improved while remaining interpretable, by using a cost-sensitive decision tree model. Our approach aims to create a fairer system by identifying more homogeneous groups of burn patients, thus ensuring a consistent cost reimbursement for the provision of similar patient care. To validate the methodology, we use data of actual burn patients. We show that the proposed approach can identify more homogeneous patient groups, enabling a more equitable and fair reimbursement. 

\section{COHORT AND DATA DESCRIPTION}
The current burn care pathway is designed to treat paediatrics separately from adults due to evidence of young age as a significant complicator of burn injury \cite{NationalBurnCareReviewCommittee2001NationalIsles}.  Further evidence was found in \cite{Onah2019} that different factors and resource requirements characterise adult and young burn patients. Thus, any developed approach should be customised for adult and child patients. This paper focuses on identifying homogeneous groups of young (<16 years old) burn patients only, by using anonymised patient-level data from all burn services in England and Wales.

The information of patients that attended a burn service is collected by clinicians and nurses, from the first contact with a burn service to rehabilitation and any late reconstruction procedure \cite{Stylianou2015a}. The dataset excludes burn injury patients who died before reaching a burn service and those with minor injuries cared for in the community or hospitals that are not specialist burn services. The data covers 2003 to 2019, comprising just over 18,000 young patients and was extracted from the international Burn Injury Database (iBID) \cite{IBID2014}. An internal iBID patient-level costing methodology \cite{Duncan2020} is applied to generate the cost of the burn patients.

\section{ANALYTICAL PIPELINE}
In healthcare, adopted models are potentially responsible for human life and, thus, there is a need for high confidence from clinicians on the results of a model. Therefore, it is imperative that adopted models are explainable. An explainable model, such as a decision tree, allows result tracing, increases transparency and facilitates model improvement \cite{Pawar2020ExplainableHealthcare}. For instance, offering a clear explanation of why a specific burn patient is placed in a higher severity group than another would increase the trust level of clinicians. Thus, we propose a cost-sensitive decision tree model, trained and tested to evaluate model performance on unseen data. 

A decision tree classifier was chosen as it allows the identification of rules generated for classification. To increase homogeneity, our paper introduces a key feature: the allocation of misclassification costs between classes. We chose a cost-sensitive decision tree that ensures that even with a misclassification, the predicted class is in close proximity (by cost, LOS and severity) to the expected patient class. This model enables model penalisation as the distance of the class predicted compared to the actual class increases \cite{Domingos1999}. This distance is identified by ranking classes by the average value of LOS, cost and total burn surface area (TBSA). The classifier was implemented with the mlr \cite{Bischl2016} and rpart \cite{Therneau2019} packages in R, with a customised cost measure.

In summary, the analytical pipeline adopted to build this proposed model includes data preprocessing, identifying target features and variables of importance, and building the final model.

\subsection{Data Preprocessing}
With the need to compare our adopted model results to the current HRG groups, access to the HRG labels for each patient is needed. However, due to non-access to the Health Episode Statistics (HES) dataset typically used to create HRGs and thus the inability to directly merge the iBID dataset to the HES, it is essential to recreate the HRG burn methodology using the iBID dataset. This was done in close collaboration with the NHS National Casemix office. The first step was to replace all missing values with 0. The assumption is that fields are left empty when the value is 0, no or not applicable (also adopted by the current HRG methodology) \cite{Keuss2018}.  The next step was the identification of iBID variables or a proxy to variables used in the HES dataset. The HRG replication was then done using if-else rules created by the Casemix team and adapted by an NHS Casemix expert to suit the iBID dataset. We expect 13 classes to be generated for the young burn patient subgroup, ranked from most severe to least severe.

The next data preprocessing step involves removing all cases grouped into the unclassifiable group, as these cases have no TBSA or depth of burn recorded for any of the 27 potential burn sites (parts of the body), which is a pre-requisite (as stated in the burn HRG methodology) for reimbursement as a burn case. Cases with LOS greater than 360 and cost greater than £1,000,000 are considered outliers by the iBID data administrators and thus removed.  The data preprocessing also includes removing irrelevant variables (administrative variables, variables with high missing values, variables with unstructured data, variables with one unique value and variables which are duplicates of another).

\subsection{Identification of Target Features and Variables of Importance}
The proposed model is a decision tree (DT) classifier, a supervised model that requires ‘true’ labels. Thus, the target classes (true labels) were generated using three critical factors in burn care – LOS, cost and TBSA. These factors were identified from previous analysis and the literature and represent the aim of HRGs to create clinically meaningful groups (burn severity) that have similar resource usage (LOS and cost of treatment). These factors were first log-transformed to reduce skew, followed by k-means cluster analysis to segment each factor into 13 target classes. The identified classes were ranked from most severe/costly to least severe/costly (to allow for adequate comparison with the current HRGs). With these, three cost-sensitive DT models (LOS, TBSA and cost DT)  were built using the ranked classes to identify the variables of importance from each model. The target class ranking is necessary to enable comparison with the ranked HRG and the incorporation of a cost-sensitive model.

\subsection{Final Decision Tree (DT) Model}
Here, we describe the final model for comparison with the current HRG and extraction of classification rules. The 13 target classes for this model are derived using k-means cluster analysis. The feature space for the cluster analysis is computed by calculating the average rank across all LOS, TBSA and cost DT target classes as identified using the approach described in Section 3.2. Adopting the mean rank of each case allows for the identification of classes that reflect the average severity, cost and length of hospital stay.  

The independent variables are those identified as important from the LOS, TBSA and cost DT. The dataset is then split into a train and test set. These subsets are oversampled independently by duplicating cases in the minority classes. The oversampling is needed to ensure the adopted model has enough samples to learn the decision boundary effectively.

The model used is a cost-sensitive DT model, trained and tested to evaluate model performance on unseen data. The cost-sensitive DT model is used to increase model penalisation as the distance of the class predicted compared to the actual class increases. 

\section{Results}
\begin{figure}[hbt!]
  \centering
  \includegraphics[width=\linewidth]{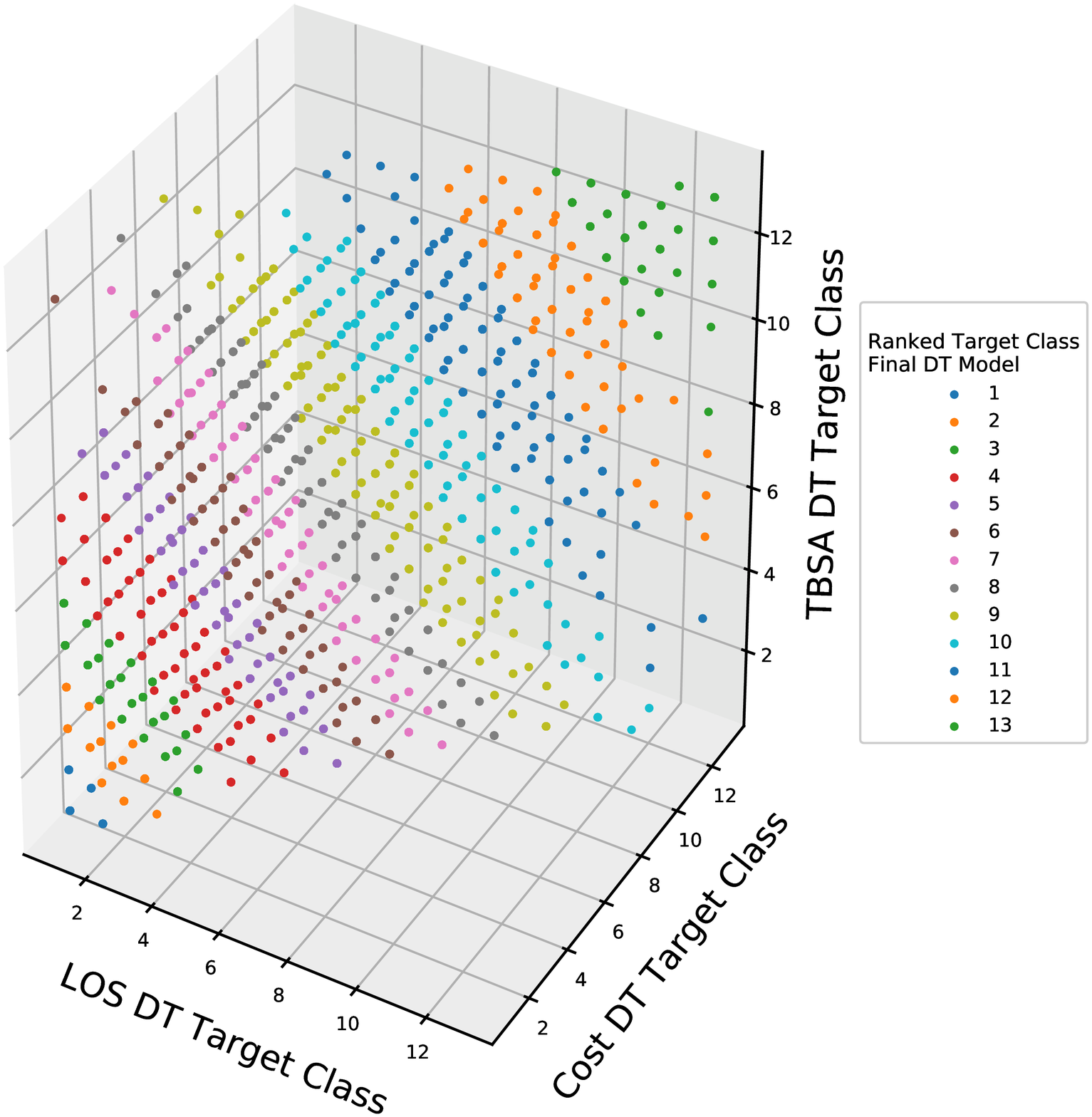}
  \caption{Spread of ranked LOS, Cost and TBSA model target classes - each dot represents a patient's LOS, cost and TBSA group and the colour represents the Final DT group.}
  \label{fig:spread}
  \Description{Shows the distribution of LOS/TBSA/Cost DT by the final DT target classes}
\end{figure}

This section presents the results of the data-driven model generated using a cost-sensitive DT. We first evaluate the engineered target classes generated to ensure it reflects homogeneity in severity and resource usage. Figure \ref{fig:spread} shows a scatterplot of LOS, TBSA, and Cost DT model ranked target classes, where class 1 depicts least severe/costly, and class 13 depicts most severe/costly, against that of the Final DT target classes (also ranked). The plotted Final DT target classes reflects, as expected, homogeneity in severity and resource usage. It shows that the target classes have a decreasing LOS, cost, and TBSA as its severity (reflected by the Final DT ranked classes) decreases.

\begin{figure}[hbt!]
  \centering
  \includegraphics[width=\linewidth]{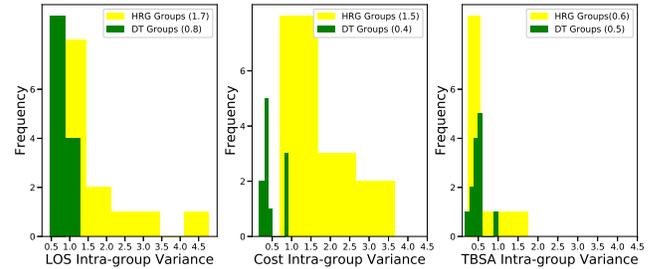}
  \caption{DT and HRG intra-group variance - LOS, Cost and TBSA - on train data. The value on each legend is the average Cost, LOS and TBSA intra-group variance for the DT and HRG groups.}
  \label{fig:variance}
  \Description{This shows the cluster variance measured on LOS, TBSA and cost.}
\end{figure}

Comparing the HRGs with the generated target classes, we show histograms of intra-group variance (in terms of cost, TBSA and LOS) of the DT and HRG model to evaluate how much homogeneity can be found.  Figure \ref{fig:variance} illustrates a lower intra-group variance in the DT groups than to the HRG groups for the train data. In particular, we found that, on average, the cost intra-group variance (0.4) for DT groups is three times lower than the HRG groups cost intra-group variance (1.5). This pattern is seen for the LOS intra-group variance - 0.8 for DT groups and 1.7 for HRG groups. As expected, a lower difference when we compare the average severity (TBSA) intra-group variance of DT groups at 0.5 and that of the HRG groups at 0.6. Evaluating these intra-group variances on the train data is needed to ensure the engineered classes meet the homogeneity criteria. These evaluations indicate that the new target classes provide higher intra-group similarity in terms of the three critical factors than the HRGs.  Thus, these target classes are adopted.

\begin{figure}[hbt!]
  \centering
  \includegraphics[width=\linewidth]{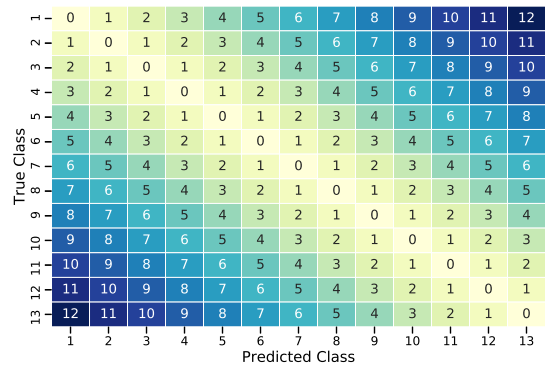}
  \caption{Cost matrix representing the penalty used in the cost-sensitive DT to ensure any misclassifications are close to the true class.}
  \label{fig:costmatrix}
  \Description{This shows the cost of misclassification}
\end{figure}

With these promising results, it is imperative to understand the rules that allow the classification of cases into each engineered class and evaluate the model’s performance on unseen cases. This is done by identifying the feature space, which are the variables of importance identified from the TBSA, LOS and Cost DT classifiers. Then the creation of a cost matrix for penalisation. The cost matrix is shown in Figure \ref{fig:costmatrix}, with the rows indicating true and the columns predicted class labels. The diagonal zero values indicate a cost of 0 for accurate prediction, with an increase in cost as predicted class deviation from the true class increases.  This is added to the Final DT model pipeline with a train and test split to evaluate performance. 

\begin{figure}[hbt!]
  \centering
  \includegraphics[width=\linewidth]{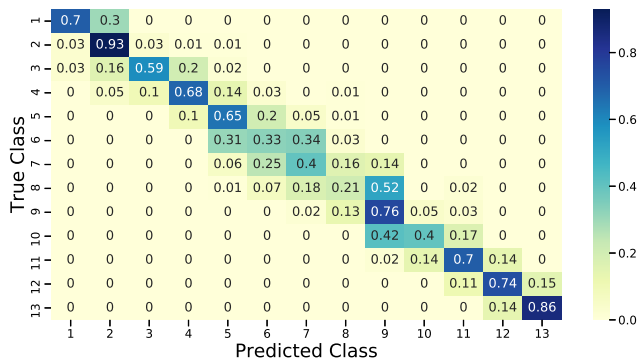}
  \caption{Confusion matrix representing the impact of penalisation on the DT model, using the test data, this shows that all misclassifications are at most 3 neighbours away from the true class.}
  \label{fig:confmat}
  \Description{This shows the cost of misclassification}
\end{figure}

\begin{figure}[hbt!]
  \centering
  \includegraphics[width=\linewidth]{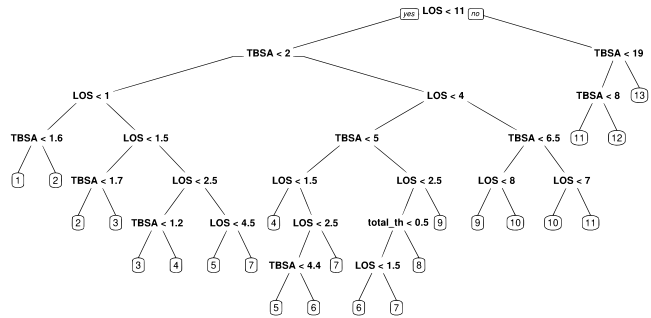}
  \caption{Classification tree showing explainable classification rules generated from the cost-sensitive DT model.}
  \label{fig:decisiontree}
  \Description{This is a view of the decision tree}
\end{figure}

The confusion matrix shown in Figure \ref{fig:confmat} reveals good classification of unseen cases (test data). The confusion matrix shows that most of the cases are predicted as the precise class. Where the classification is wrong, the misclassified class is in close proximity to the true class. Except for class 8, with most cases predicted as class 9 and some cases misclassified three classes away.  This higher misclassification of class 8 suggests the likely need to reduce the total number of classes in the young patient subgroup.  The model identifies class 8 and 9 as both made up of cases with TBSA >=5 but differentiated by the LOS. Class 8 are those with LOS < 2.5 while class 9 are patients with LOS >= 2.5 (see Figure \ref{fig:decisiontree}).

\begin{figure}[hbt!]
  \centering
  \includegraphics[width=\linewidth]{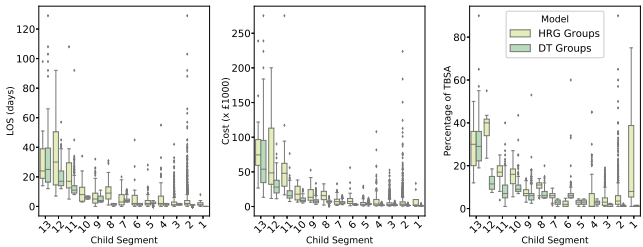}
  \caption{DT and HRG groups by LOS, Cost and TBSA on test data. Ordered in decreasing order of injury severity.}
  \label{fig:boxplot}
  \Description{The image shows a boxplot of critical factor distribution in the HRG and DT groups.}
\end{figure}
\begin{figure}[hbt!]
  \centering
  \includegraphics[width=\linewidth]{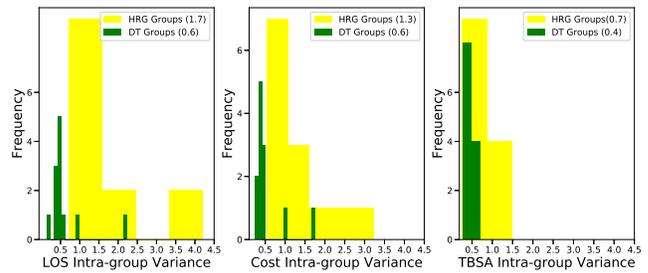}
  \caption{DT and HRG intra-group variance - LOS, Cost and TBSA - on test data. The value on each legend is the average LOS, Cost and TBSA intra-group variance for the DT and HRG groups.}
  \label{fig:varianceTest}
  \Description{This shows the predicted DT versus HRG intra-group variance measured on LOS, TBSA and cost.}
\end{figure}

Figure \ref{fig:boxplot} illustrates the difference in the homogeneity of the HRG groups and the DT groups using a boxplot to show the distribution of TBSA, LOS and cost in each group. The groups are ordered by severity/resource usage, with class 13 representing HRG and DT groups with the most severe cases and high resource usage. Class 1 represents the groups with the least severe and least costly cases in both the HRG and DT groups. In both, as expected, we see a gradual reduction in these three factors as severity and resource usage decreases. Notably, this is more apparent in the DT groups, which also have shorter boxplots indicating higher homogeneity across DT groups. Comparing the cluster variance (using LOS, Cost and TBSA) of the predicted groups to the HRG groups of the same cases, we confirm this higher homogeneity. As seen in Figure \ref{fig:varianceTest} on average, the DT groups have a lower intra-group variance compared to the HRG groups.

Exploring the classification rules as shown in Figure \ref{fig:decisiontree}, the adopted model does well in producing an explainable result. The classification tree illustrates the criteria for adding a case to a particular group. Class 1's (the least severe/costly group) case membership are those with LOS < 1 and TBSA < 1.6 while class 13 (the most severe/costly group) are made up of cases with LOS >=11 and TBSA >= 19. We can also identify the three most important features as LOS, TBSA, and total theatre visits from the classification tree.

\section{Conclusion and Future Work}

The adopted model creates a classification rule that is easy to explain and less complex than the current HRG methodology. This meets the HRG requirement that groups are clinically relevant to allow for organisational activities. Additional, our adopted model meets the requirements of group homogeneity in terms of resource usage and severity. The resultant groups from the adopted DT model evidenced increased homogeneity across the three critical factors – length of hospital stay, cost to the hospital and severity of burn, compared to the current HRG. Thus, our identified groups are better suited for cost reimbursement. 

Our future work includes the customisation of the cost matrix to reflect a non-linear and non-monotonic penalty (e.g. actual monetary cost) for misclassification and applying this analytical pipeline to adult burn patients.


\bibliographystyle{ACM-Reference-Format}

\end{document}